\documentclass[%
reprint,
superscriptaddress,
 amsmath,amssymb,
pra,
]{revtex4-1}

\usepackage{graphicx}
\usepackage{dcolumn}
\usepackage{bm}

\begin{document}


\title{Low Gilbert damping in Co$_2$FeSi and Fe$_2$CoSi films}%

\author{Christian Sterwerf}%
\email{csterwerf@physik.uni-bielefeld.de}%
\affiliation{Center for Spinelectronic Materials and Devices, Physics Department, Bielefeld University, Germany}%
\author{Soumalya Paul}%
\affiliation{Department of Physics and Astronomy/MINT Center, The University of Alabama, Tuscaloosa, AL 35487, USA}%
\author{Behrouz Khodadadi}%
\affiliation{Department of Physics and Astronomy/MINT Center, The University of Alabama, Tuscaloosa, AL 35487, USA}%
\author{Markus~Meinert}%
\affiliation{Center for Spinelectronic Materials and Devices, Physics Department, Bielefeld University, Germany}%
\author{Jan-Michael~Schmalhorst}%
\affiliation{Center for Spinelectronic Materials and Devices, Physics Department, Bielefeld University, Germany}%
\author{Mathias Buchmeier}%
\affiliation{Department of Physics and Astronomy/MINT Center, The University of Alabama, Tuscaloosa, AL 35487, USA}%
\author{Claudia K. A. Mewes}%
\affiliation{Department of Physics and Astronomy/MINT Center, The University of Alabama, Tuscaloosa, AL 35487, USA}%
\author{Tim Mewes}%
\affiliation{Department of Physics and Astronomy/MINT Center, The University of Alabama, Tuscaloosa, AL 35487, USA}%
\author{G\"unter~Reiss}%
\affiliation{Center for Spinelectronic Materials and Devices, Physics Department, Bielefeld University, Germany}%

\date{\today}

\begin{abstract}
Thin highly textured Fe$_{\mathrm{1+x}}$Co$_{\mathrm{2-x}}$Si ($0 \leq$ x $\leq 1$) films were prepared on MgO (001) substrates by magnetron co-sputtering. The magneto-optic Kerr effect (MOKE) and ferromagnetic resonance (FMR) measurements were used to investigate the composition dependence of the magnetization, the magnetic anisotropy, the gyromagnetic ratio and the relaxation of the films.
The effective magnetization for the thin Fe$_{\mathrm{1+x}}$Co$_{\mathrm{2-x}}$Si films, determined by FMR measurements, are consistent with the Slater Pauling prediction. Both MOKE and FMR measurements reveal a pronounced fourfold anisotropy distribution for all films. In addition we found a strong influence of the stoichiometry on the anisotropy as the cubic anisotropy strongly increases with increasing Fe concentration. The gyromagnetic ratio is only weakly dependent on the composition. We find low Gilbert damping parameters for all films with values down to $0.0012\pm0.00012$ for Fe$_{1.75}$Co$_{1.25}$Si. The effective damping parameter for Co$_2$FeSi is found to be $0.0018\pm 0.0004$.
We also find a pronounced anisotropic relaxation, which indicates significant contributions of two-magnon scattering processes that is strongest along the easy axes of the films.
This makes thin Fe$_{\mathrm{1+x}}$Co$_{\mathrm{2-x}}$Si films ideal materials for the application in STT-MRAM devices.
\end{abstract}

\pacs{Valid PACS appear here}
\maketitle


\section{Introduction}
Half-metallic ferromagnets have attracted great interest during the past few years because they promise to boost the performance of spintronic devices. High spin polarization at the Fermi level can generate high tunnel magnetoresistance (TMR) ratios. A TMR effect can be measured in a magnetic tunnel junction (MTJ) that consists of two ferromagnetic films separated by a thin insulator. The same structures can also be utilized to spin transfer torque induced magnetization switching \cite{Berger:1996}, however in this case a low switching current density is desirable. Thus, low magnetic damping and a high spin polarization are frequently required for spin transfer torque based devices \cite{Slonczewski:1996}. A high spin polarization can be found in half-metals where one spin band structure is semiconducting while the other spin band structure is metallic. Co- and Fe-based Heusler compounds are good candidates for materials with high Curie temperatures and half-metallic behavior.\\
Full Heusler compounds have the formula X$_2$YZ, where X and Y are transition metals and Z is a main group element. There are two different ordered structures: the L$2_1$ structure and the X$_{\mathrm{a}}$ structure with a different occupation sequence. Both structures consist of a four-atom basis and an fcc lattice. The prototype of the L$2_1$ structure is Cu$_2$MnAl (space group Fm$\bar{3}$m) with the occupation sequence X-Y-X-Z \cite{Bradley1934}. The prototypes for the X$_{\mathrm{a}}$ structure are Hg$_2$CuTi and Li$_2$AgSb with an occupation sequence Y-X-X-Z, with the two X-atoms at inequivalent positions in the lattice \cite{puselj,pauly}.
In this work, we investigate the magnetic properties of a stoichiometric series ranging from Co$_2$FeSi to Fe$_2$CoSi, where Co$_2$FeSi crystalizes in the L$2_1$ structure and Fe$_2$CoSi in the X$_{\mathrm{a}}$ structure, respectively. Both compounds should have a (pseudo-)gap in the minority states as predicted by first principle calculations. By substituting Co and Fe atoms the number of electrons varies and the Fermi level is expected to be shifted to lower energies when the Fe concentration is increased. As we reported previously, magnetic tunnel junctions based on the Fe$_{\mathrm{1+x}}$Co$_{\mathrm{2-x}}$Si films exhibit very high TMR ratios for all stoichiometries \cite{Sterwerf:2013}. At $15$\,K a maximum TMR ratio of $262$\% was found for the intermediate stoichiometry Fe$_{1.75}$Co$_{1.25}$Si, while the Co$_2$FeSi and Fe$_2$CoSi based MTJs showed a TMR ratio of $167$\% and $227$\%, respectively. One possible explanation for the high TMR ratio is that for Fe$_{1.75}$Co$_{1.25}$Si the Fermi energy is shifted inside the pseudo-gap. In this work we present results of the magnetic properties for the magnetization dynamics in particular including anisotropy and the Gilbert damping parameter of the Fe$_{\mathrm{1+x}}$Co$_{\mathrm{2-x}}$Si films, as the intrinsic relaxation is are expected to be low for half-metals \cite{Liu:2009dm}.

\section{Preparation and Characterization Techniques}
Thin Fe$_{\mathrm{1+x}}$Co$_{\mathrm{2-x}}$Si (x=$0$, $0.25$, $0.5$, $0.75$, $1$) films were fabricated using co-sputtering in an UHV sputtering system with a base pressure of $1\cdot 10^{-9}$\,mbar. The Ar pressure during sputtering was $2\cdot 10^{-3}$\,mbar. The films were grown by dc- and rf-magnetron sputtering from elemental targets onto MgO ($001$) substrates. Additional MgO and Cr seed layers were used to accommodate small lattice mismatches and to promote coherent and epitaxial growth, as the Cr seed layer grows in $45^\circ$ direction on the MgO layer, which has a lattice parameter of $4.212$\,\AA{}. The lattice mismatch between two unit cells of Cr ($2\times2.885$\,\AA{} at $20^\circ$C \cite{Straumanis:a01440}) and one unit cell of Co$_2$FeSi ($5.64$\,\AA{} \cite{Wurmehl:2005ia}) or Fe$_2$CoSi ($5.645$\,\AA{} \cite{Luo:2007io}) is about $2$\%. The $5$\,nm thick MgO and Cr films were in-situ annealed at $700^\circ$C to obtain smooth surfaces. Fe$_{\mathrm{1+x}}$Co$_{\mathrm{2-x}}$Si films with a thickness of $20$\,nm were deposited at room temperature and ex-situ vacuum annealed at $500^\circ$C. A $2$\,nm thick MgO capping layer was used to prevent oxidation of the films. To determine the stoichiometry and to adjust the sputtering powers, x-ray fluorescence measurements were carried out. 
To obtain information about the magnetization dynamics, in-plane ferromagnetic resonance (FMR) measurements were performed using a broadband coplanar waveguide setup up to a maximum frequency of $40$\,GHz. Least square fits of the raw data using a first derivative of a Lorenzian line shape were done to precisely determine the resonance field and the peak-to-peak linewidth $\Delta H$ \cite{Schabes:2000,Pachauri:2015ia}.
For the FMR in-plane angle dependent measurements the samples were mounted on a rotating stage and the resonance spectra were measured at a frequency of $30$\,GHz while the in-plane angle was changed in $5^\circ$ steps. 
In addition quasistatic magnetization reversal measurements were carried out using the magneto-optic Kerr effect (MOKE) in a vector MOKE setup with an s-polarized laser with a wavelength of $488$\,nm. Anisotropy measurements were carried out using a rotating sample holder. The magnetic field was applied in the plane of the films.

\section{Crystallographic Properties}
\begin{figure}[!t]%
\centering%
\includegraphics[width=\linewidth]{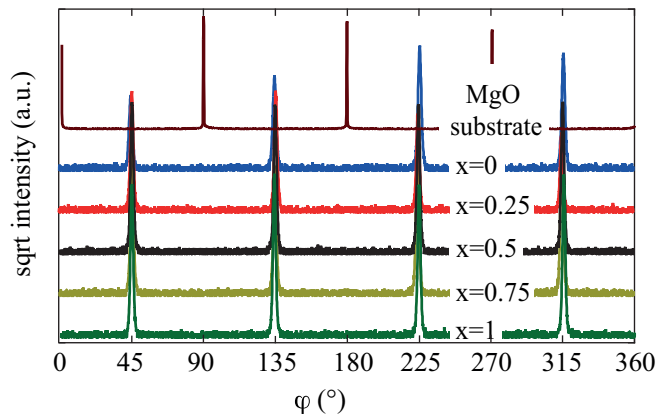}%
\caption{$\varphi$-scans of the ($220$) Fe$_{\mathrm{1+x}}$Co$_{\mathrm{2-x}}$Si peak and ($220$) MgO substrate peak showing the fourfold symmetry of the films.}%
\label{phi-scans}%
\end{figure}

X-ray diffraction measurements were used to investigate the crystallographic properties of the Fe$_{\mathrm{1+x}}$Co$_{\mathrm{2-x}}$Si films. Ordering parameters, determined from x-ray diffraction, were already discussed in our previous work \cite{Sterwerf:2013} and found to be high for Co$_2$FeSi and decrease when going to Fe$_2$CoSi. In order to test the films for crystallographic symmetry $\varphi$ scans are performed on the ($220$) planes of the Fe$_{\mathrm{1+x}}$Co$_{\mathrm{2-x}}$Si films. Figure \ref{phi-scans} shows the results together with the ($220$) plane of the MgO ($001$) substrate. The result shows that the ($100$) Heusler plane is rotated by $45^\circ$ with respect to the MgO ($100$) plane. The fourfold symmetry of the $\varphi$-scans clearly verifies the highly textured growth of all Fe$_{\mathrm{1+x}}$Co$_{\mathrm{2-x}}$Si films of this study.

\section{Magnetization dynamics}
\begin{figure}[!t]%
\centering%
\includegraphics[width=\linewidth]{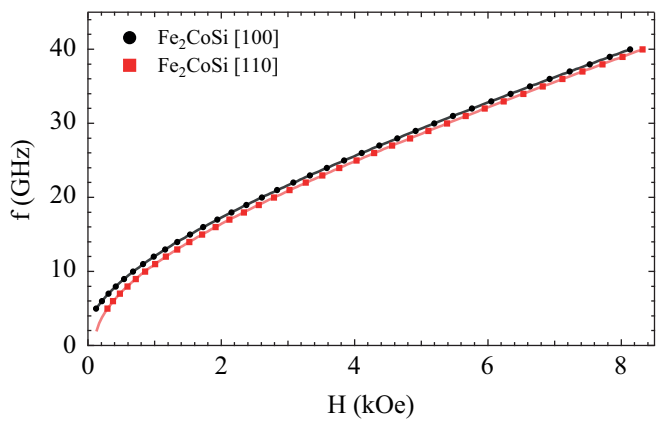}%
\caption{Resonance frequency versus magnetic field (Kittel plot) along the in-plane magnetic hard [$110$] and the magnetic easy [$100$] axis for Fe$_2$CoSi. The experimental data are fitted using a combined fit (equations (\ref{Kittelformel1} and \ref{Kittelformel})) to determine $M_{\mathrm{eff}}$ and $\gamma'$.}%
\label{Fe2CoSi_Kittel}%
\end{figure}%
In this section we present in-plane broadband FMR measurements for the Fe$_{\mathrm{1+x}}$Co$_{\mathrm{2-x}}$Si samples to obtain information about the magnetic properties of the films. The Landau-Lifshitz-Gilbert equation describes the dynamics of the magnetization vector $\vec{M}$ in the presence of an effective field $\vec{H}_{\mathrm{eff}}$, which contains both dc and ac fields.

\begin{figure}[!t]%
\centering%
\includegraphics[width=0.8\linewidth]{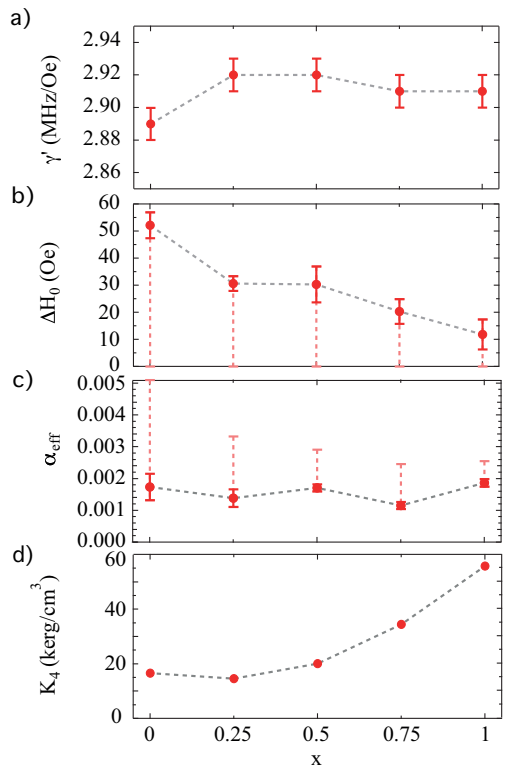}%
\caption{a) Gyromagnetic ratio $\gamma'$, b) Extrinsic contribution to the linewidth $\Delta H_0$ of the FMR spectra, c) effective Gilbert damping parameter and d) cubic magnetic anisotropy constant $K_4$ for Fe$_{\mathrm{1+x}}$Co$_{\mathrm{2-x}}$Si films with $x=0$, $0.25$, $0.5$, $0.75$ ,$1$.}%
\label{a_dH0_vs_x}%
\end{figure}

It is given by \cite{Heinrich:1993}:
\begin{eqnarray}
\frac{\mathrm{d} \vec{M}}{\mathrm{d} t}=-\gamma\vec{M}\times\vec{H}_{\mathrm{eff}}+\frac{\alpha}{M}\left(\vec{M}\times\frac{\mathrm{d}\vec{M}}{\mathrm{d}t}\right),
\label{LLG}
\end{eqnarray}
where $\gamma$ is the gyromagnetic ratio and the quantity parameter $\alpha$ is the Gilbert damping parameter.
According to the Landau-Lifshitz-Gilbert equation (\ref{LLG}), the resonance condition can be expressed in terms of the second derivatives of the free-energy density $E$ by the Smit-Beljers formula \cite{Suhl:1955}:
\begin{eqnarray}
\left(\frac{f}{\gamma'}\right)^2=\left.\frac{1}{(M \sin{\theta})^2}\left[\frac{\partial ^2 E}{\partial \theta^2}\frac{\partial ^2 E}{\partial \varphi^2}-\left(\frac{\partial ^2 E}{\partial \theta\partial\varphi}\right)^2 \right]\right|_{\theta_0,\varphi_0},
\label{Suhl}
\end{eqnarray}
where $\gamma'=\gamma/2\pi$, $\theta$ and $\varphi$ are the polar and azimuthal angles of the magnetization $\vec{M}$ and $\theta_0$ and $\varphi_0$ the corresponding equilibrium values.
Measurements of the magnetic field dependent resonance frequency were carried out in two different orientations of the sample: in [$100$] and [$110$] direction of the Fe$_{\mathrm{1+x}}$Co$_{\mathrm{2-x}}$Si Heusler alloy, as the [$100$] direction is the magnetic easy axis and the [$110$] direction the magnetic hard axis, respectively. Figure \ref{Fe2CoSi_Kittel} shows the exemplary Kittel plots along [$100$] and [$110$] directions for the Fe$_2$CoSi sample. The experimental data were fitted simultaneously using the Kittel equation for both easy and hard configurations \cite{Liu:2003ca}:
\begin{eqnarray}
f&=&\gamma' \sqrt{(H_{\mathrm{res-ha}}-H_4)(H_{\mathrm{res-ha}}+\frac{H_4}{2}+4\pi M_{\mathrm{eff}})}\label{Kittelformel1}\\
f&=&\gamma' \sqrt{(H_{\mathrm{res-ea}}+H_4)(H_{\mathrm{res-ea}}+H_4+4\pi M_{\mathrm{eff}})}
\label{Kittelformel}
\end{eqnarray}
where $M_{\mathrm{eff}}$, $\gamma'$ and $H_4$ are shared fit parameters. $H_4$ describes the magnitude of the in-plane fourfold anisotropy field. $H_{\mathrm{res-ha}}$ and $H_{\mathrm{res-ea}}$ denote the resonance field along the magnetic hard and the magnetic easy axis, respectively. The resulting fit parameters for the gyromagnetic ratio $\gamma'$ are presented in Fig. \ref{a_dH0_vs_x} a) for all $x$ in Fe$_{\mathrm{1+x}}$Co$_{\mathrm{2-x}}$Si. Within the error bars it is nearly constant for x $\geq$ $0.25$ and slightly smaller for Co$_2$FeSi.
\begin{figure}[!t]%
\centering%
\includegraphics[width=\linewidth]{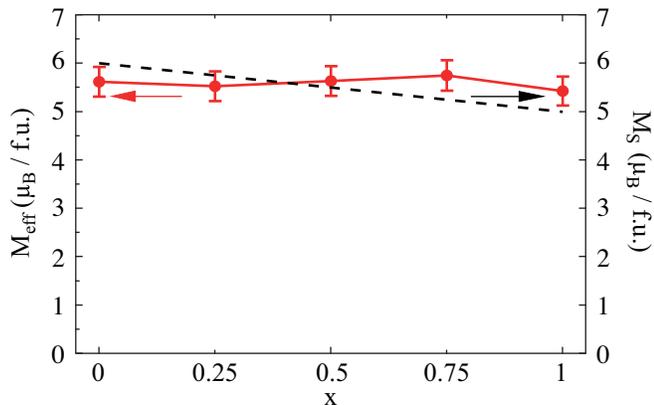}%
\caption{Dependence of the fitted effective magnetic moment per formula unit for Fe$_{\mathrm{1+x}}$Co$_{\mathrm{2-x}}$Si films with x=$0$, $0.25$, $0.5$, $0.75$, $1$ shown on the left axis. The dashed line shows the interpolated expected magnetic moments according to the Slater-Pauling rule (right axis).}%
\label{magnetization_vs_x}%
\end{figure}%
The fitted effective magnetization, which includes any perpendicular anisotropy present in the films, is shown in Fig. \ref{magnetization_vs_x} for the Fe$_{\mathrm{1+x}}$Co$_{\mathrm{2-x}}$Si samples. The error bars originate from fitting of the Kittel equations and the determination of the volume of the unit cell. For bulk Co$_2$FeSi and Fe$_2$CoSi the experimentally determined magnetizations are $5.95\,\mu_{\mathrm{B}}$/f.u.\cite{Wurmehl:2005ia} and $4.99\,\mu_{\mathrm{B}}$/f.u.\cite{Luo:2007io}, respectively, which match the expected magnetizations according to the Slater-Pauling rule (visualized by the dashed line in Fig. \ref{magnetization_vs_x} on the right axis). The deviation from the expected values might be attributed to residual atomic disorder in the films or the presence of a perpendicular anisotropy caused by a small tetragonal distortion in the [$001$] direction.

\begin{figure}[!t]%
\centering%
\includegraphics[width=\linewidth]{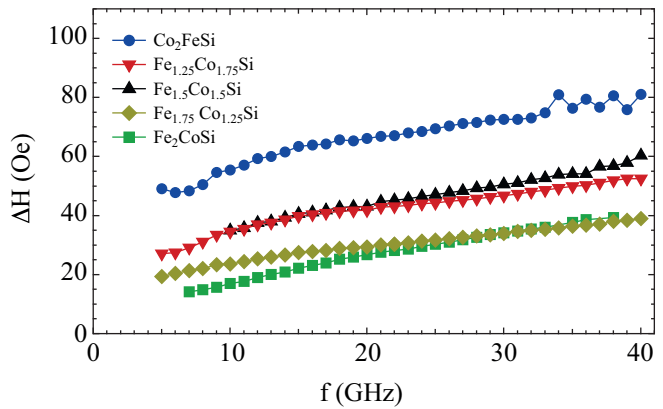}%
\caption{Frequency dependent FMR linewidth for all samples measured along the magnetic hard axis [$110$] of the Fe$_{\mathrm{1+x}}$Co$_{\mathrm{2-x}}$Si films.}%
\label{dH_vs_f}%
\end{figure}

The frequency dependence of the linewidth of the ferromagnetic resonance absorption provides direct information about the magnetic relaxation. The frequency dependence of the linewidth \cite{Heinrich:1985,Lee:2008gz} can under certain conditions be characterized by an inhomogeneous residual linewidth at zero field $\Delta H_0$ and an intrinsic contribution \cite{Mewesbuch}:
\begin{eqnarray}
\Delta H=\Delta H_0 + \frac{2}{\sqrt{3}}\frac{\alpha_{\mathrm{eff}}}{\gamma'} f.
\label{linewidthformel}
\end{eqnarray}
For correct determination of the effective damping parameter it is necessary to measure the linewidth over a wide frequency range to determine the slope. It is not sufficient to measure $\Delta H$ at a fixed frequency, because a non-zero extrinsic linewidth $\Delta H_0$ results in an overestimated damping parameter $\alpha_{\mathrm{eff}}$.
Figure \ref{dH_vs_f} shows the peak-to-peak linewidth $\Delta H$ for all frequencies and all x. The measurements were performed in the direction of the magnetic hard axis of the Heusler films. The experimental data were fitted by equation (\ref{linewidthformel}) to determine the effective damping parameters. The slope at higher frequencies was used to determine the damping parameters.
The inhomogeneous residual linewidth at zero field $\Delta H_0$ is presented in Fig. \ref{a_dH0_vs_x} b) for all stoichiometries. The error margins result from the different slopes in the $\Delta H$ vs. $f$ curves. The residual linewidth decreases as the Fe concentration increases and reaches its lowest value of $\Delta H_0=12$\,Oe for Fe$_2$CoSi. McMichael \textit{et al.} \cite{PhysRevLett.90.227601} found that small grain size distributions can lead to low inhomogeneous line broadening. 

The effective Gilbert damping parameter $\alpha_{\mathrm{eff}}$ is shown in Fig. \ref{a_dH0_vs_x} c). All damping parameters have the same order of magnitude and vary between $0.0012\pm0.00012$ to $0.0019\pm0.00013$. The very upper limit of the error margins was calculated assuming that the linewidth measured at $40$\,GHz is caused solely by Gilbert type damping. Co$_2$FeSi exhibits a damping parameter of $0.0018\pm 0.0004$, while Fe$_2$CoSi shows a slightly larger value of $0.0019\pm 0.00013$.  Kasatani \textit{et al.} found damping parameters from $0.0023$ to $0.0061$ for Co$_2$FeSi films and $0.002$ for Fe$_2$CoSi \cite{Kasatani:2014}. In general, the Gilbert damping is expected to be low in half-metallic materials, where spin-flip processes are suppressed \cite{Muller:2009vr,Liu:2009dm}. The small damping parameters of the metallic films show that a pseudo-gap as present in the Fe$_{\mathrm{1+x}}$Co$_{\mathrm{2-x}}$Si system is sufficient to give rise to a low Gilbert damping.

\begin{figure}[!t]%
\centering%
\includegraphics[width=\linewidth]{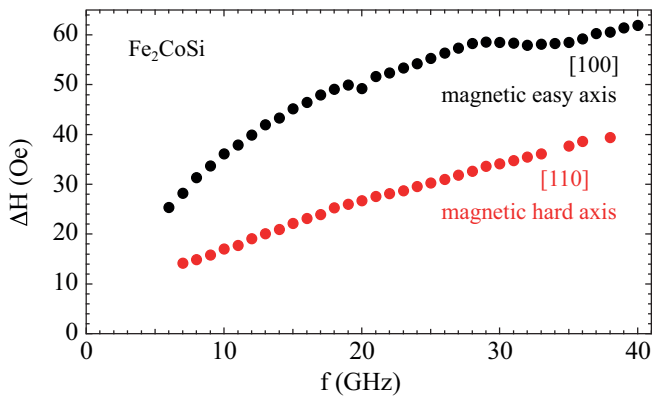}%
\caption{FMR linewidth for Fe$_2$CoSi measured along both the magnetic hard [$110$] and magnetic easy [$100$] axis.}%
\label{dH_vs_f_Fe2CoSi}%
\end{figure}%

Figure \ref{dH_vs_f_Fe2CoSi} shows the frequency dependent linewidth along easy and hard axes for the Fe$_2$CoSi. The linewidth exhibits almost linear behavior (the Gilbert model) along the hard axis. We observed non-linear behavior in the linewidth vs. frequency response along the magnetic easy axis. This nonlinear dependence of the FMR linewidth on frequency is a typical observation when two magnon scattering contributes significantly to the relaxation \cite{Lee:2009gk,Landeros:2008fm}. Two-magnon scattering is an extrinsic relaxation mechanism and can be induced by means of different scattering centers such as voids or pores \cite{Hurben:1998bb}, surface roughness \cite{Lee:2009gk} and grain size \cite{McMichael:1998bs} or by network of misfit dislocations which causes scattering of the FMR mode (k=0) into propagating spin waves (k$\neq$0).

\begin{figure}[!t]%
\centering%
\includegraphics[width=1\linewidth]{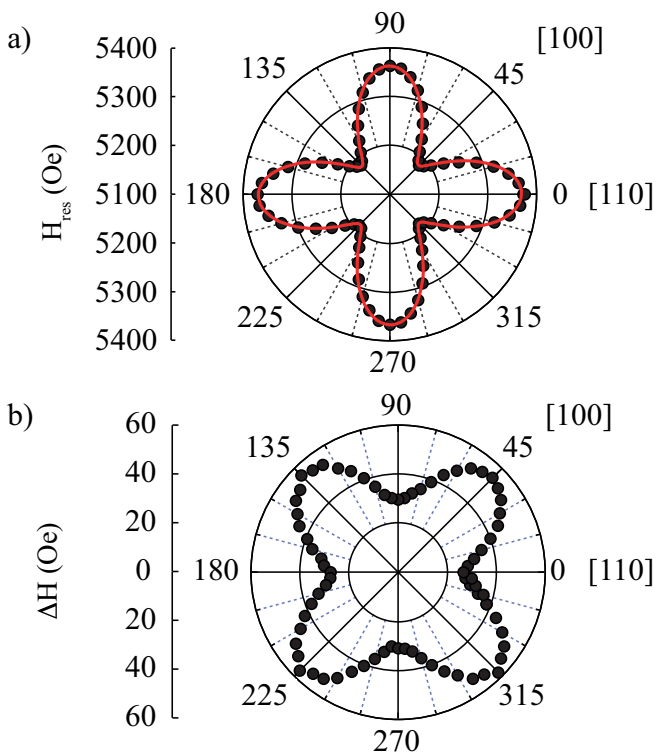}%
\caption{Polar plots of a) the resonance fields H$_{\mathrm{res}}$ and b) the linewidth $\Delta H$ as a function of the in-plane angle of the applied field with respect to the [$110$] axis of a $20$\,nm thick Fe$_2$CoSi film measured at a microwave frequency of $30$\,GHz.}%
\label{H_res_polar}%
\end{figure}%

\begin{figure}[!t]%
\centering%
\includegraphics[width=1\linewidth]{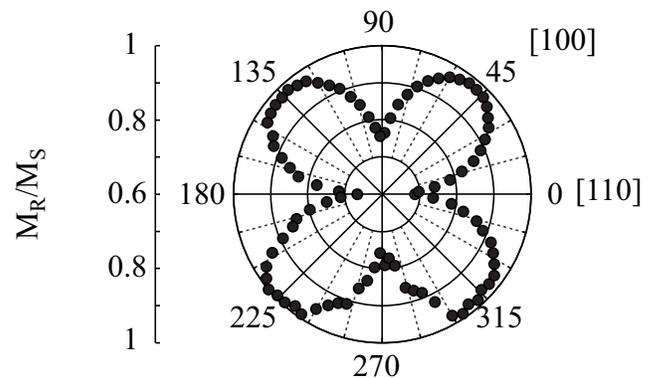}%
\caption{Polar plots of the squareness $\frac{M_{\mathrm{R}}}{M_{\mathrm{S}}}$ for Fe$_2$CoSi obtained by MOKE measurements.}%
\label{Fe2CoSi_MOKE}%
\end{figure}%
\subsection{FMR in-plane rotation measurements}
To obtain further information about the magnetic anisotropies and magnetic relaxation additional FMR measurements were carried out as a function of the in-plane angle of the applied field with respect to the Fe$_{\mathrm{1+x}}$Co$_{\mathrm{2-x}}$Si [$110$] axis. The operating frequency for the rotation measurements was $30$\,GHz. At this frequency the resonance fields are high enough to saturate the magnetization along the easy and hard axes. All measurements were performed at room temperature.

A fourfold symmetry is observed in the in-plane angle dependence of the ferromagnetic resonance field for all samples. Figure \ref{H_res_polar} a) exemplarily shows the ferromagnetic resonance field $H_{\mathrm{res}}$ versus the in-plane rotation angle for Fe$_2$CoSi. The dependence of the resonance field on the in-plane angle was simulated numerically using equation (\ref{Suhl}), assuming a cubic magnetic anisotropy contribution to the Gibbs free energy \cite{Farle:1998uy,HeinrichBuch}:
\begin{eqnarray}
E_{\mathrm{cubic}}=-\frac{1}{2}K_4\left(\alpha_1^4+\alpha_2^4+\alpha_3^4 \right),
\label{E4}
\end{eqnarray}
where $K_4$ is the cubic magnetic anisotropy constant and $\alpha_1$, $\alpha_2$, $\alpha_3$ are the directional cosines with respect to the cubic principal axes. The experimentally determined in-plane angle dependent $H_{\mathrm{res}}$ data were fitted with the numerical solution (red line in Fig. \ref{H_res_polar} a)) to determine the cubic anisotropy constant. 
Figure \ref{H_res_polar} b) shows the corresponding linewidth data, which also shows a clear fourfold symmetry. The linewidth exhibits maxima along the easy axes and minima along the hard axes of the cubic magnetic anisotropy. Randomly distributed crystalline defects oriented along the in-plane principal crystallographic axis \cite{PhysRevB.85.014420} or a fourfold distribution in misfit dislocations \cite{Woltersdorf:2004hh} which induce the same symmetry on the strength of two magnon scattering can explain the observed anisotropic relaxation.

The magnetic fourfold symmetry matches the crystallographic symmetry of the highly textured Fe$_{\mathrm{1+x}}$Co$_{\mathrm{2-x}}$Si films mentioned before.
A polar plot of the MOKE squareness versus the rotational angle for Fe$_2$CoSi is presented in Fig. \ref{Fe2CoSi_MOKE}. This measurement confirms the cubic anisotropy present in the films as seen in the FMR measurement. The magnetic easy axis is located along the [$100$] crystallographic axis and the magnetic hard axis is located along the [$110$] crystallographic axis. A cubic anisotropy with the easy magnetic axis in the Heusler [$100$] direction is found for all samples. The cubic magnetic anisotropy constant $K_4$ obtained from the FMR measurements changes significantly in this series from $55.8\,\frac{\mathrm{kerg}}{\mathrm{cm}^3}$ for Fe$_2$CoSi to $16.6\,\frac{\mathrm{kerg}}{\mathrm{cm}^3}$ for Co$_2$FeSi, respectively. The cubic anisotropy constants for all stoichiometries are presented in Fig. \ref{a_dH0_vs_x} d). Hashimoto \textit{et al.} found a similar cubic anisotropy constant of $18\,\frac{\mathrm{kerg}}{\mathrm{cm}^3}$ for crystalline Co$_2$FeSi with a film thickness of $18.5$\,nm \cite{Hashimoto:2005de}. Some films show an additional uniaxial anisotropy component, which can originate from miscut substrates.

\section{Conclusion}
In summary we found very small damping parameters for the half-metallic Fe$_{\mathrm{1+x}}$Co$_{\mathrm{2-x}}$Si films varying from $0.0012\pm0.00012$ to $0.0019\pm0.00013$. Co$_2$FeSi exhibits a damping parameter of $0.0018\pm 0.0004$. Thus, the films are suitable for the use in STT-MRAMs.
FMR and MOKE measurements reveal a fourfold magnetocrystalline anisotropy for all films in accordance with the fourfold crystalline symmetry in the highly textured films. The need for frequency dependent FMR measurements was exemplified by the finding that the residual linewidth changes both with composition and with the measurement direction.

\begin{acknowledgments}
The authors gratefully acknowledge financial support from Bundesministerium f\"ur Bildung und Forschung (BMBF) and Deutsche Forschungsgemeinschaft (DFG, contract no. RE 1052/32-1) as well as support through the MINT Center summer program. S. Paul, B. Khodadadi and T. Mewes would like to acknowledge support by the NSF-CAREER Award No. 0952929, C.K.A. Mewes would like to acknowledge support by the NSF-CAREER Award No. 1452670.
\end{acknowledgments}

\bibliography{Fe2CoSi.bib}

\begin{thebibliography}{30}%
\makeatletter
\providecommand \@ifxundefined [1]{%
 \@ifx{#1\undefined}
}%
\providecommand \@ifnum [1]{%
 \ifnum #1\expandafter \@firstoftwo
 \else \expandafter \@secondoftwo
 \fi
}%
\providecommand \@ifx [1]{%
 \ifx #1\expandafter \@firstoftwo
 \else \expandafter \@secondoftwo
 \fi
}%
\providecommand \natexlab [1]{#1}%
\providecommand \enquote  [1]{``#1''}%
\providecommand \bibnamefont  [1]{#1}%
\providecommand \bibfnamefont [1]{#1}%
\providecommand \citenamefont [1]{#1}%
\providecommand \href@noop [0]{\@secondoftwo}%
\providecommand \href [0]{\begingroup \@sanitize@url \@href}%
\providecommand \@href[1]{\@@startlink{#1}\@@href}%
\providecommand \@@href[1]{\endgroup#1\@@endlink}%
\providecommand \@sanitize@url [0]{\catcode `\\12\catcode `\$12\catcode
  `\&12\catcode `\#12\catcode `\^12\catcode `\_12\catcode `\%12\relax}%
\providecommand \@@startlink[1]{}%
\providecommand \@@endlink[0]{}%
\providecommand \url  [0]{\begingroup\@sanitize@url \@url }%
\providecommand \@url [1]{\endgroup\@href {#1}{\urlprefix }}%
\providecommand \urlprefix  [0]{URL }%
\providecommand \Eprint [0]{\href }%
\providecommand \doibase [0]{http://dx.doi.org/}%
\providecommand \selectlanguage [0]{\@gobble}%
\providecommand \bibinfo  [0]{\@secondoftwo}%
\providecommand \bibfield  [0]{\@secondoftwo}%
\providecommand \translation [1]{[#1]}%
\providecommand \BibitemOpen [0]{}%
\providecommand \bibitemStop [0]{}%
\providecommand \bibitemNoStop [0]{.\EOS\space}%
\providecommand \EOS [0]{\spacefactor3000\relax}%
\providecommand \BibitemShut  [1]{\csname bibitem#1\endcsname}%
\let\auto@bib@innerbib\@empty
\bibitem [{\citenamefont {Berger}(1996)}]{Berger:1996}%
  \BibitemOpen
  \bibfield  {author} {\bibinfo {author} {\bibfnamefont {L.}~\bibnamefont
  {Berger}},\ }\href@noop {} {\bibfield  {journal} {\bibinfo  {journal}
  {Physical Review B}\ }\textbf {\bibinfo {volume} {54}},\ \bibinfo {pages}
  {9353} (\bibinfo {year} {1996})}\BibitemShut {NoStop}%
\bibitem [{\citenamefont {Slonczewski}(1996)}]{Slonczewski:1996}%
  \BibitemOpen
  \bibfield  {author} {\bibinfo {author} {\bibfnamefont {J.~C.}\ \bibnamefont
  {Slonczewski}},\ }\href@noop {} {\bibfield  {journal} {\bibinfo  {journal}
  {Journal of Magnetism and Magnetic Materials}\ }\textbf {\bibinfo {volume}
  {159}},\ \bibinfo {pages} {L1} (\bibinfo {year} {1996})}\BibitemShut
  {NoStop}%
\bibitem [{\citenamefont {Bradley}\ and\ \citenamefont
  {Rodgers}(1934)}]{Bradley1934}%
  \BibitemOpen
  \bibfield  {author} {\bibinfo {author} {\bibfnamefont {A.~J.}\ \bibnamefont
  {Bradley}}\ and\ \bibinfo {author} {\bibfnamefont {J.~W.}\ \bibnamefont
  {Rodgers}},\ }\href@noop {} {\bibfield  {journal} {\bibinfo  {journal}
  {Proceedings of the Royal Society of London Series A}\ }\textbf {\bibinfo
  {volume} {144}},\ \bibinfo {pages} {340} (\bibinfo {year}
  {1934})}\BibitemShut {NoStop}%
\bibitem [{\citenamefont {Puselj}\ and\ \citenamefont {Ban}(1969)}]{puselj}%
  \BibitemOpen
  \bibfield  {author} {\bibinfo {author} {\bibfnamefont {M.}~\bibnamefont
  {Puselj}}\ and\ \bibinfo {author} {\bibfnamefont {Z.}~\bibnamefont {Ban}},\
  }\href@noop {} {\bibfield  {journal} {\bibinfo  {journal} {Croat. Chem.
  Acta}\ }\textbf {\bibinfo {volume} {41}},\ \bibinfo {pages} {79} (\bibinfo
  {year} {1969})}\BibitemShut {NoStop}%
\bibitem [{\citenamefont {Pauly}\ \emph {et~al.}(1968)\citenamefont {Pauly},
  \citenamefont {Weiss},\ and\ \citenamefont {Witte}}]{pauly}%
  \BibitemOpen
  \bibfield  {author} {\bibinfo {author} {\bibfnamefont {H.}~\bibnamefont
  {Pauly}}, \bibinfo {author} {\bibfnamefont {A.}~\bibnamefont {Weiss}}, \ and\
  \bibinfo {author} {\bibfnamefont {H.}~\bibnamefont {Witte}},\ }\href@noop {}
  {\bibfield  {journal} {\bibinfo  {journal} {Z. Metallkunde}\ }\textbf
  {\bibinfo {volume} {59}},\ \bibinfo {pages} {47} (\bibinfo {year}
  {1968})}\BibitemShut {NoStop}%
\bibitem [{\citenamefont {Sterwerf}\ \emph {et~al.}(2013)\citenamefont
  {Sterwerf}, \citenamefont {Meinert}, \citenamefont {Schmalhorst},\ and\
  \citenamefont {Reiss}}]{Sterwerf:2013}%
  \BibitemOpen
  \bibfield  {author} {\bibinfo {author} {\bibfnamefont {C.}~\bibnamefont
  {Sterwerf}}, \bibinfo {author} {\bibfnamefont {M.}~\bibnamefont {Meinert}},
  \bibinfo {author} {\bibfnamefont {J.-M.}\ \bibnamefont {Schmalhorst}}, \ and\
  \bibinfo {author} {\bibfnamefont {G.}~\bibnamefont {Reiss}},\ }\href@noop {}
  {\bibfield  {journal} {\bibinfo  {journal} {IEEE Transactions on Magnetics}\
  }\textbf {\bibinfo {volume} {49}},\ \bibinfo {pages} {4386} (\bibinfo {year}
  {2013})}\BibitemShut {NoStop}%
\bibitem [{\citenamefont {Liu}\ \emph {et~al.}(2009)\citenamefont {Liu},
  \citenamefont {Mewes}, \citenamefont {Chshiev},\ and\ \citenamefont
  {Mewes}}]{Liu:2009dm}%
  \BibitemOpen
  \bibfield  {author} {\bibinfo {author} {\bibfnamefont {C.}~\bibnamefont
  {Liu}}, \bibinfo {author} {\bibfnamefont {C.}~\bibnamefont {Mewes}}, \bibinfo
  {author} {\bibfnamefont {M.}~\bibnamefont {Chshiev}}, \ and\ \bibinfo
  {author} {\bibfnamefont {T.}~\bibnamefont {Mewes}},\ }\href@noop {}
  {\bibfield  {journal} {\bibinfo  {journal} {Applied Physics Letters}\
  }\textbf {\bibinfo {volume} {95}},\ \bibinfo {pages} {022509} (\bibinfo
  {year} {2009})}\BibitemShut {NoStop}%
\bibitem [{\citenamefont {Straumanis}\ and\ \citenamefont
  {Weng}(1955)}]{Straumanis:a01440}%
  \BibitemOpen
  \bibfield  {author} {\bibinfo {author} {\bibfnamefont {M.~E.}\ \bibnamefont
  {Straumanis}}\ and\ \bibinfo {author} {\bibfnamefont {C.~C.}\ \bibnamefont
  {Weng}},\ }\href {\doibase 10.1107/S0365110X55001254} {\bibfield  {journal}
  {\bibinfo  {journal} {Acta Crystallographica}\ }\textbf {\bibinfo {volume}
  {8}},\ \bibinfo {pages} {367} (\bibinfo {year} {1955})}\BibitemShut {NoStop}%
\bibitem [{\citenamefont {Wurmehl}\ \emph {et~al.}(2005)\citenamefont
  {Wurmehl}, \citenamefont {Fecher}, \citenamefont {Kandpal}, \citenamefont
  {Ksenofontov}, \citenamefont {Felser}, \citenamefont {Lin},\ and\
  \citenamefont {Morais}}]{Wurmehl:2005ia}%
  \BibitemOpen
  \bibfield  {author} {\bibinfo {author} {\bibfnamefont {S.}~\bibnamefont
  {Wurmehl}}, \bibinfo {author} {\bibfnamefont {G.}~\bibnamefont {Fecher}},
  \bibinfo {author} {\bibfnamefont {H.}~\bibnamefont {Kandpal}}, \bibinfo
  {author} {\bibfnamefont {V.}~\bibnamefont {Ksenofontov}}, \bibinfo {author}
  {\bibfnamefont {C.}~\bibnamefont {Felser}}, \bibinfo {author} {\bibfnamefont
  {H.-J.}\ \bibnamefont {Lin}}, \ and\ \bibinfo {author} {\bibfnamefont
  {J.}~\bibnamefont {Morais}},\ }\href@noop {} {\bibfield  {journal} {\bibinfo
  {journal} {Physical Review B}\ }\textbf {\bibinfo {volume} {72}},\ \bibinfo
  {pages} {184434} (\bibinfo {year} {2005})}\BibitemShut {NoStop}%
\bibitem [{\citenamefont {Luo}\ \emph {et~al.}(2007)\citenamefont {Luo},
  \citenamefont {Zhu}, \citenamefont {Ma}, \citenamefont {Xu}, \citenamefont
  {Liu}, \citenamefont {Qu}, \citenamefont {Li},\ and\ \citenamefont
  {Wu}}]{Luo:2007io}%
  \BibitemOpen
  \bibfield  {author} {\bibinfo {author} {\bibfnamefont {H.}~\bibnamefont
  {Luo}}, \bibinfo {author} {\bibfnamefont {Z.}~\bibnamefont {Zhu}}, \bibinfo
  {author} {\bibfnamefont {L.}~\bibnamefont {Ma}}, \bibinfo {author}
  {\bibfnamefont {S.}~\bibnamefont {Xu}}, \bibinfo {author} {\bibfnamefont
  {H.}~\bibnamefont {Liu}}, \bibinfo {author} {\bibfnamefont {J.}~\bibnamefont
  {Qu}}, \bibinfo {author} {\bibfnamefont {Y.}~\bibnamefont {Li}}, \ and\
  \bibinfo {author} {\bibfnamefont {G.}~\bibnamefont {Wu}},\ }\href@noop {}
  {\bibfield  {journal} {\bibinfo  {journal} {Journal of Physics D: Applied
  Physics}\ }\textbf {\bibinfo {volume} {40}},\ \bibinfo {pages} {7121}
  (\bibinfo {year} {2007})}\BibitemShut {NoStop}%
\bibitem [{\citenamefont {Schabes}\ \emph {et~al.}(2000)\citenamefont
  {Schabes}, \citenamefont {Zhou},\ and\ \citenamefont
  {Bertram}}]{Schabes:2000}%
  \BibitemOpen
  \bibfield  {author} {\bibinfo {author} {\bibfnamefont {M.~E.}\ \bibnamefont
  {Schabes}}, \bibinfo {author} {\bibfnamefont {H.}~\bibnamefont {Zhou}}, \
  and\ \bibinfo {author} {\bibfnamefont {H.~N.}\ \bibnamefont {Bertram}},\
  }\href@noop {} {\bibfield  {journal} {\bibinfo  {journal} {Journal of Applied
  Physics}\ }\textbf {\bibinfo {volume} {87}},\ \bibinfo {pages} {5666}
  (\bibinfo {year} {2000})}\BibitemShut {NoStop}%
\bibitem [{\citenamefont {Pachauri}\ \emph {et~al.}(2015)\citenamefont
  {Pachauri}, \citenamefont {Khodadadi},\ and\ \citenamefont
  {Althammer}}]{Pachauri:2015ia}%
  \BibitemOpen
  \bibfield  {author} {\bibinfo {author} {\bibfnamefont {N.}~\bibnamefont
  {Pachauri}}, \bibinfo {author} {\bibfnamefont {B.}~\bibnamefont {Khodadadi}},
  \ and\ \bibinfo {author} {\bibfnamefont {M.}~\bibnamefont {Althammer}},\
  }\href@noop {} {\bibfield  {journal} {\bibinfo  {journal} {Journal of Applied
  Physics}\ }\textbf {\bibinfo {volume} {117}},\ \bibinfo {pages} {233907}
  (\bibinfo {year} {2015})}\BibitemShut {NoStop}%
\bibitem [{\citenamefont {Heinrich}\ and\ \citenamefont
  {Cochran}(1993)}]{Heinrich:1993}%
  \BibitemOpen
  \bibfield  {author} {\bibinfo {author} {\bibfnamefont {B.}~\bibnamefont
  {Heinrich}}\ and\ \bibinfo {author} {\bibfnamefont {J.~F.}\ \bibnamefont
  {Cochran}},\ }\href@noop {} {\bibfield  {journal} {\bibinfo  {journal}
  {Advances in Physics}\ }\textbf {\bibinfo {volume} {42}},\ \bibinfo {pages}
  {523} (\bibinfo {year} {1993})}\BibitemShut {NoStop}%
\bibitem [{\citenamefont {Suhl}(1955)}]{Suhl:1955}%
  \BibitemOpen
  \bibfield  {author} {\bibinfo {author} {\bibfnamefont {H.}~\bibnamefont
  {Suhl}},\ }\href@noop {} {\bibfield  {journal} {\bibinfo  {journal} {Physical
  Review}\ }\textbf {\bibinfo {volume} {97}},\ \bibinfo {pages} {555} (\bibinfo
  {year} {1955})}\BibitemShut {NoStop}%
\bibitem [{\citenamefont {Liu}\ \emph {et~al.}(2003)\citenamefont {Liu},
  \citenamefont {Sasaki},\ and\ \citenamefont {Furdyna}}]{Liu:2003ca}%
  \BibitemOpen
  \bibfield  {author} {\bibinfo {author} {\bibfnamefont {X.}~\bibnamefont
  {Liu}}, \bibinfo {author} {\bibfnamefont {Y.}~\bibnamefont {Sasaki}}, \ and\
  \bibinfo {author} {\bibfnamefont {J.~K.}\ \bibnamefont {Furdyna}},\
  }\href@noop {} {\bibfield  {journal} {\bibinfo  {journal} {Physical Review
  B}\ }\textbf {\bibinfo {volume} {67}},\ \bibinfo {pages} {205204} (\bibinfo
  {year} {2003})}\BibitemShut {NoStop}%
\bibitem [{\citenamefont {Heinrich}\ \emph {et~al.}(1985)\citenamefont
  {Heinrich}, \citenamefont {Cochran},\ and\ \citenamefont
  {Hasegawa}}]{Heinrich:1985}%
  \BibitemOpen
  \bibfield  {author} {\bibinfo {author} {\bibfnamefont {B.}~\bibnamefont
  {Heinrich}}, \bibinfo {author} {\bibfnamefont {J.~F.}\ \bibnamefont
  {Cochran}}, \ and\ \bibinfo {author} {\bibfnamefont {R.}~\bibnamefont
  {Hasegawa}},\ }\href@noop {} {\bibfield  {journal} {\bibinfo  {journal}
  {Journal of Applied Physics}\ }\textbf {\bibinfo {volume} {57}},\ \bibinfo
  {pages} {3690} (\bibinfo {year} {1985})}\BibitemShut {NoStop}%
\bibitem [{\citenamefont {Lee}\ \emph {et~al.}(2008)\citenamefont {Lee},
  \citenamefont {Wen}, \citenamefont {Pathak},\ and\ \citenamefont
  {Janssen}}]{Lee:2008gz}%
  \BibitemOpen
  \bibfield  {author} {\bibinfo {author} {\bibfnamefont {H.}~\bibnamefont
  {Lee}}, \bibinfo {author} {\bibfnamefont {L.}~\bibnamefont {Wen}}, \bibinfo
  {author} {\bibfnamefont {M.}~\bibnamefont {Pathak}}, \ and\ \bibinfo {author}
  {\bibfnamefont {P.}~\bibnamefont {Janssen}},\ }\href@noop {} {\bibfield
  {journal} {\bibinfo  {journal} {Journal of Physics D: Applied Physics}\
  }\textbf {\bibinfo {volume} {41}},\ \bibinfo {pages} {215001} (\bibinfo
  {year} {2008})}\BibitemShut {NoStop}%
\bibitem [{\citenamefont {Mewes}\ and\ \citenamefont
  {Mewes}(2015)}]{Mewesbuch}%
  \BibitemOpen
  \bibinfo {editor} {\bibfnamefont {C.}~\bibnamefont {Mewes}}\ and\ \bibinfo
  {editor} {\bibfnamefont {T.}~\bibnamefont {Mewes}},\ \href@noop {}
  {\emph {\bibinfo {title} {{Relaxation in Magnetic Materials for Spintronics,
  in: Handbook of Nanomagnetism}}}}\ (\bibinfo  {publisher} {Pan Stanford},\
  \bibinfo {year} {2015})\ p.~\bibinfo {pages} {74}\BibitemShut {NoStop}%
\bibitem [{\citenamefont {McMichael}\ \emph {et~al.}(2003)\citenamefont
  {McMichael}, \citenamefont {Twisselmann},\ and\ \citenamefont
  {Kunz}}]{PhysRevLett.90.227601}%
  \BibitemOpen
  \bibfield  {author} {\bibinfo {author} {\bibfnamefont {R.~D.}\ \bibnamefont
  {McMichael}}, \bibinfo {author} {\bibfnamefont {D.~J.}\ \bibnamefont
  {Twisselmann}}, \ and\ \bibinfo {author} {\bibfnamefont {A.}~\bibnamefont
  {Kunz}},\ }\href {\doibase 10.1103/PhysRevLett.90.227601} {\bibfield
  {journal} {\bibinfo  {journal} {Phys. Rev. Lett.}\ }\textbf {\bibinfo
  {volume} {90}},\ \bibinfo {pages} {227601} (\bibinfo {year}
  {2003})}\BibitemShut {NoStop}%
\bibitem [{\citenamefont {Kasatani}\ \emph {et~al.}(2014)\citenamefont
  {Kasatani}, \citenamefont {Yamada}, \citenamefont {Itoh}, \citenamefont
  {Miyao}, \citenamefont {Hamaya},\ and\ \citenamefont
  {Nozaki}}]{Kasatani:2014}%
  \BibitemOpen
  \bibfield  {author} {\bibinfo {author} {\bibfnamefont {Y.}~\bibnamefont
  {Kasatani}}, \bibinfo {author} {\bibfnamefont {S.}~\bibnamefont {Yamada}},
  \bibinfo {author} {\bibfnamefont {H.}~\bibnamefont {Itoh}}, \bibinfo {author}
  {\bibfnamefont {M.}~\bibnamefont {Miyao}}, \bibinfo {author} {\bibfnamefont
  {K.}~\bibnamefont {Hamaya}}, \ and\ \bibinfo {author} {\bibfnamefont
  {Y.}~\bibnamefont {Nozaki}},\ }\href@noop {} {\bibfield  {journal} {\bibinfo
  {journal} {Applied Physics Express}\ }\textbf {\bibinfo {volume} {7}},\
  \bibinfo {pages} {123001} (\bibinfo {year} {2014})}\BibitemShut {NoStop}%
\bibitem [{\citenamefont {M{\"u}ller}\ \emph {et~al.}(2009)\citenamefont
  {M{\"u}ller}, \citenamefont {Walowski}, \citenamefont {Djordjevic},\ and\
  \citenamefont {Miao}}]{Muller:2009vr}%
  \BibitemOpen
  \bibfield  {author} {\bibinfo {author} {\bibfnamefont {G.~M.}\ \bibnamefont
  {M{\"u}ller}}, \bibinfo {author} {\bibfnamefont {J.}~\bibnamefont
  {Walowski}}, \bibinfo {author} {\bibfnamefont {M.}~\bibnamefont
  {Djordjevic}}, \ and\ \bibinfo {author} {\bibfnamefont {G.~X.}\ \bibnamefont
  {Miao}},\ }\href@noop {} {\bibfield  {journal} {\bibinfo  {journal} {Nature
  Materials}\ }\textbf {\bibinfo {volume} {8}},\ \bibinfo {pages} {56}
  (\bibinfo {year} {2009})}\BibitemShut {NoStop}%
\bibitem [{\citenamefont {Lee}\ \emph {et~al.}(2009)\citenamefont {Lee},
  \citenamefont {Wang}, \citenamefont {Mewes},\ and\ \citenamefont
  {Butler}}]{Lee:2009gk}%
  \BibitemOpen
  \bibfield  {author} {\bibinfo {author} {\bibfnamefont {H.}~\bibnamefont
  {Lee}}, \bibinfo {author} {\bibfnamefont {Y.}~\bibnamefont {Wang}}, \bibinfo
  {author} {\bibfnamefont {C.}~\bibnamefont {Mewes}}, \ and\ \bibinfo {author}
  {\bibfnamefont {W.~H.}\ \bibnamefont {Butler}},\ }\href@noop {} {\bibfield
  {journal} {\bibinfo  {journal} {Applied Physics Letters}\ }\textbf {\bibinfo
  {volume} {95}},\ \bibinfo {pages} {082502} (\bibinfo {year}
  {2009})}\BibitemShut {NoStop}%
\bibitem [{\citenamefont {Landeros}\ \emph {et~al.}(2008)\citenamefont
  {Landeros}, \citenamefont {Arias},\ and\ \citenamefont
  {Mills}}]{Landeros:2008fm}%
  \BibitemOpen
  \bibfield  {author} {\bibinfo {author} {\bibfnamefont {P.}~\bibnamefont
  {Landeros}}, \bibinfo {author} {\bibfnamefont {R.~E.}\ \bibnamefont {Arias}},
  \ and\ \bibinfo {author} {\bibfnamefont {D.~L.}\ \bibnamefont {Mills}},\
  }\href@noop {} {\bibfield  {journal} {\bibinfo  {journal} {Physical Review
  B}\ }\textbf {\bibinfo {volume} {77}},\ \bibinfo {pages} {214405} (\bibinfo
  {year} {2008})}\BibitemShut {NoStop}%
\bibitem [{\citenamefont {Hurben}\ and\ \citenamefont
  {Patton}(1998)}]{Hurben:1998bb}%
  \BibitemOpen
  \bibfield  {author} {\bibinfo {author} {\bibfnamefont {M.~J.}\ \bibnamefont
  {Hurben}}\ and\ \bibinfo {author} {\bibfnamefont {C.~E.}\ \bibnamefont
  {Patton}},\ }\href@noop {} {\bibfield  {journal} {\bibinfo  {journal}
  {Journal of Applied Physics}\ }\textbf {\bibinfo {volume} {83}},\ \bibinfo
  {pages} {4344} (\bibinfo {year} {1998})}\BibitemShut {NoStop}%
\bibitem [{\citenamefont {McMichael}\ \emph {et~al.}(1998)\citenamefont
  {McMichael}, \citenamefont {Stiles},\ and\ \citenamefont
  {Chen}}]{McMichael:1998bs}%
  \BibitemOpen
  \bibfield  {author} {\bibinfo {author} {\bibfnamefont {R.~D.}\ \bibnamefont
  {McMichael}}, \bibinfo {author} {\bibfnamefont {M.~D.}\ \bibnamefont
  {Stiles}}, \ and\ \bibinfo {author} {\bibfnamefont {P.~J.}\ \bibnamefont
  {Chen}},\ }\href@noop {} {\bibfield  {journal} {\bibinfo  {journal} {Journal
  of Applied Physics}\ }\textbf {\bibinfo {volume} {83}},\ \bibinfo {pages}
  {7037} (\bibinfo {year} {1998})}\BibitemShut {NoStop}%
\bibitem [{\citenamefont {Farle}(1998)}]{Farle:1998uy}%
  \BibitemOpen
  \bibfield  {author} {\bibinfo {author} {\bibfnamefont {M.}~\bibnamefont
  {Farle}},\ }\href@noop {} {\bibfield  {journal} {\bibinfo  {journal} {Reports
  on Progress in Physics}\ }\textbf {\bibinfo {volume} {61}},\ \bibinfo {pages}
  {755} (\bibinfo {year} {1998})}\BibitemShut {NoStop}%
\bibitem [{\citenamefont {Heinrich}\ and\ \citenamefont
  {Bland}(1994)}]{HeinrichBuch}%
  \BibitemOpen
  \bibinfo {editor} {\bibfnamefont {B.}~\bibnamefont {Heinrich}}\ and\ \bibinfo
  {editor} {\bibfnamefont {J.}~\bibnamefont {Bland}},\ eds.,\ \href@noop {}
  {\emph {\bibinfo {title} {{Radio Frequency Techniques, in: Ultrathin Magnetic
  Structures II}}}}\ (\bibinfo  {publisher} {Springer},\ \bibinfo {year}
  {1994})\ p.\ \bibinfo {pages} {195}\BibitemShut {NoStop}%
\bibitem [{\citenamefont {Barsukov}\ \emph {et~al.}(2012)\citenamefont
  {Barsukov}, \citenamefont {Landeros}, \citenamefont {Meckenstock},
  \citenamefont {Lindner}, \citenamefont {Spoddig}, \citenamefont {Li},
  \citenamefont {Krumme}, \citenamefont {Wende}, \citenamefont {Mills},\ and\
  \citenamefont {Farle}}]{PhysRevB.85.014420}%
  \BibitemOpen
  \bibfield  {author} {\bibinfo {author} {\bibfnamefont {I.}~\bibnamefont
  {Barsukov}}, \bibinfo {author} {\bibfnamefont {P.}~\bibnamefont {Landeros}},
  \bibinfo {author} {\bibfnamefont {R.}~\bibnamefont {Meckenstock}}, \bibinfo
  {author} {\bibfnamefont {J.}~\bibnamefont {Lindner}}, \bibinfo {author}
  {\bibfnamefont {D.}~\bibnamefont {Spoddig}}, \bibinfo {author} {\bibfnamefont
  {Z.-A.}\ \bibnamefont {Li}}, \bibinfo {author} {\bibfnamefont
  {B.}~\bibnamefont {Krumme}}, \bibinfo {author} {\bibfnamefont
  {H.}~\bibnamefont {Wende}}, \bibinfo {author} {\bibfnamefont {D.~L.}\
  \bibnamefont {Mills}}, \ and\ \bibinfo {author} {\bibfnamefont
  {M.}~\bibnamefont {Farle}},\ }\href {\doibase 10.1103/PhysRevB.85.014420}
  {\bibfield  {journal} {\bibinfo  {journal} {Phys. Rev. B}\ }\textbf {\bibinfo
  {volume} {85}},\ \bibinfo {pages} {014420} (\bibinfo {year}
  {2012})}\BibitemShut {NoStop}%
\bibitem [{\citenamefont {Woltersdorf}\ and\ \citenamefont
  {Heinrich}(2004)}]{Woltersdorf:2004hh}%
  \BibitemOpen
  \bibfield  {author} {\bibinfo {author} {\bibfnamefont {G.}~\bibnamefont
  {Woltersdorf}}\ and\ \bibinfo {author} {\bibfnamefont {B.}~\bibnamefont
  {Heinrich}},\ }\href@noop {} {\bibfield  {journal} {\bibinfo  {journal}
  {Physical Review B}\ }\textbf {\bibinfo {volume} {69}},\ \bibinfo {pages}
  {184417} (\bibinfo {year} {2004})}\BibitemShut {NoStop}%
\bibitem [{\citenamefont {Hashimoto}\ \emph {et~al.}(2005)\citenamefont
  {Hashimoto}, \citenamefont {Herfort}, \citenamefont {Schonherr},\ and\
  \citenamefont {Ploog}}]{Hashimoto:2005de}%
  \BibitemOpen
  \bibfield  {author} {\bibinfo {author} {\bibfnamefont {M.}~\bibnamefont
  {Hashimoto}}, \bibinfo {author} {\bibfnamefont {J.}~\bibnamefont {Herfort}},
  \bibinfo {author} {\bibfnamefont {H.~P.}\ \bibnamefont {Schonherr}}, \ and\
  \bibinfo {author} {\bibfnamefont {K.~H.}\ \bibnamefont {Ploog}},\ }\href@noop
  {} {\bibfield  {journal} {\bibinfo  {journal} {Applied Physics Letters}\
  }\textbf {\bibinfo {volume} {87}},\ \bibinfo {pages} {102506} (\bibinfo
  {year} {2005})}\BibitemShut {NoStop}%
\end{thebibliography}%

\end{document}